\documentclass[twocolumn,prl,superscriptaddress,floats,aps,amsmath,amssymb]{revtex4-1}
\usepackage[]{graphicx}
\usepackage{color}
\usepackage{bm} 
\usepackage{prettyref}
\usepackage{float}
\usepackage{natbib}
\usepackage{tabularx}
\usepackage{booktabs}

\newcolumntype{C}[1]{>{\centering}m{#1}}

\usepackage{array}

\begin{document}
	
	\title{Spin resonance in the superconducting state of Li$_{1-x}$Fe$_{x}$ODFe$_{1-y}$Se observed by neutron spectroscopy}
	\date{\today}
	\author{N. R. Davies}
	\author{M. C. Rahn}
	\affiliation{Department of Physics, University of Oxford, Clarendon Laboratory, Oxford, OX1 3PU, U.K.}
	\author{H. C. Walker}
	\affiliation{ISIS Facility, STFC Rutherford Appleton Laboratory, Harwell Campus, Didcot OX11 0QX, U.K.}
	\author{R. A. Ewings}
	\affiliation{ISIS Facility, STFC Rutherford Appleton Laboratory, Harwell Campus, Didcot OX11 0QX, U.K.}
	\author{D. N. Woodruff}
	\author{S. J. Clarke}
	\affiliation{Department of Chemistry, Inorganic Chemistry Laboratory, University of Oxford, South Parks Road, Oxford, OX1 3QR, U.K.}
	\author{A. T. Boothroyd}
	\email{a.boothroyd@physics.ox.ac.uk}
	\affiliation{Department of Physics, University of Oxford, Clarendon Laboratory, Oxford, OX1 3PU, U.K.}

\begin{abstract}
We have performed inelastic neutron scattering measurements on a powder sample of the superconductor lithium iron selenide hydroxide Li$_{1-x}$Fe$_{x}$ODFe$_{1-y}$Se ($x \simeq 0.16, y \simeq 0.02$, $T_{\rm c} = 41$\,K).  The spectrum shows an enhanced intensity below $T_{\rm c}$ over an energy range $0.64\times2\Delta < E < 2\Delta$, where $\Delta$ is the superconducting gap, with maxima at the wave vectors $Q_1 \simeq 1.46$\,\AA$^{-1}$ and $Q_2 \simeq 1.97$\,\AA$^{-1}$. The behavior of this feature is consistent with the spin resonance mode found in other unconventional superconductors, and strongly resembles the spin resonance observed in the spectrum of another molecular-intercalated iron selenide, Li$_{0.6}$(ND$_{2}$)$_{0.2}$(ND$_{3}$)$_{0.8}$Fe$_{2}$Se$_{2}$. The signal can be described with a characteristic two-dimensional wave vector $(\pi, 0.67\pi)$ in the Brillouin zone of the iron square lattice, consistent with the nesting vector between electron Fermi sheets.
\end{abstract}

\maketitle

\section{Introduction}

Among the iron-based superconductors, those containing FeSe layers display a particularly rich phenomenology, much of which remains unexplained \cite{Mizuguchi2010,Wen2011,Liu2015,Wu2015,Si2016}. The parent phase $\beta$-Fe$_{1+x}$Se has a relatively low superconducting transition temperature $T_{\rm c}$ of $8.5$\,K \cite{Hsu2008,McQueen2009} which, however, can be enhanced through the application of pressure \cite{Medvedev2009} or the intercalation of alkali metal ions and small molecules \cite{Guo2010,Wang2011a,Krzton2011,Fang2011,Ying2012,Burrard2013,Krzton2012,Scheidt2012}. Remarkably, superconductivity has been observed in monolayers of FeSe on SrTiO$_3$ with $T_{\rm c}$ up to 65\,K \cite{Wang2012}, and perhaps as high as $T_{\rm c} \sim 100$\,K \cite{Ge2014}.  This suggests that bulk superconductivity at similarly high temperatures might be achievable in derivatives of FeSe that have been chemically tuned to optimal carrier doping and inter-layer separation.

Separation of FeSe layers by alkali ions in compounds with bulk compositions close to $A_{0.8}$Fe$_{1.6}$Se ($A$ = K, Rb, Cs) can increase $T_{\rm c}$ up to 45\,K \cite{Guo2010,Wang2011a,Krzton2011,Fang2011}. The product, however, is inhomogeneous with a majority non-superconducting phase containing iron vacancies, and superconductivity in a minority phase \cite{Texier2012,Speller2012}. An alternative route is to synthesize intercalates of FeSe at room temperature or below from solutions of electropositive metal ions in ammonia \cite{Ying2012,Burrard2013,Krzton2012,Scheidt2012}. This method can yield single-phase material with vacancy-free FeSe layers and a controllable electronic doping level.

Recently, a new intercalated FeSe-derived bulk superconductor Li$_{1-x}$Fe$_{x}$OHFe$_{1-y}$Se has been reported with $T_{\rm c}$ in excess of 40\,K \cite{Lu2015,Pachmayr2015,Dong2014}. This material was initially synthesized by a hydrothermal route which was subsequently adapted to include a post-synthetic lithiation step, resulting in almost vacancy-free FeSe layers \cite{Sun2015}. Investigations throughout the stable composition range ($x \simeq 0.2$, $0.02 < y < 0.15$) found the highest $T_{\rm c}$ values  when the iron vacancy concentration is low $(y < 0.05)$, corresponding to significant electron-doping of the FeSe layers \cite{Sun2015}. Consistent with this, angle-resolved photoemission spectroscopy (ARPES) \cite{Niu2015,Zhao2016} and scanning tunnelling spectroscopy \cite{Du2016} measurements have shown that the Fermi surface consists only of electron pockets, centered on the X points of the iron square lattice, and that the band structure, Fermi surface and gap symmetry are very similar to those of the monolayer FeSe/SrTiO$_3$ superconductor \cite{Zhao2016}. This makes Li$_{1-x}$Fe$_{x}$OHFe$_{1-y}$Se a particularly important material in the quest to understand why certain iron-based superconductors have such high $T_{\rm c}$ values.
\begin{figure*}
	\centering
	\includegraphics[width=0.9\textwidth]{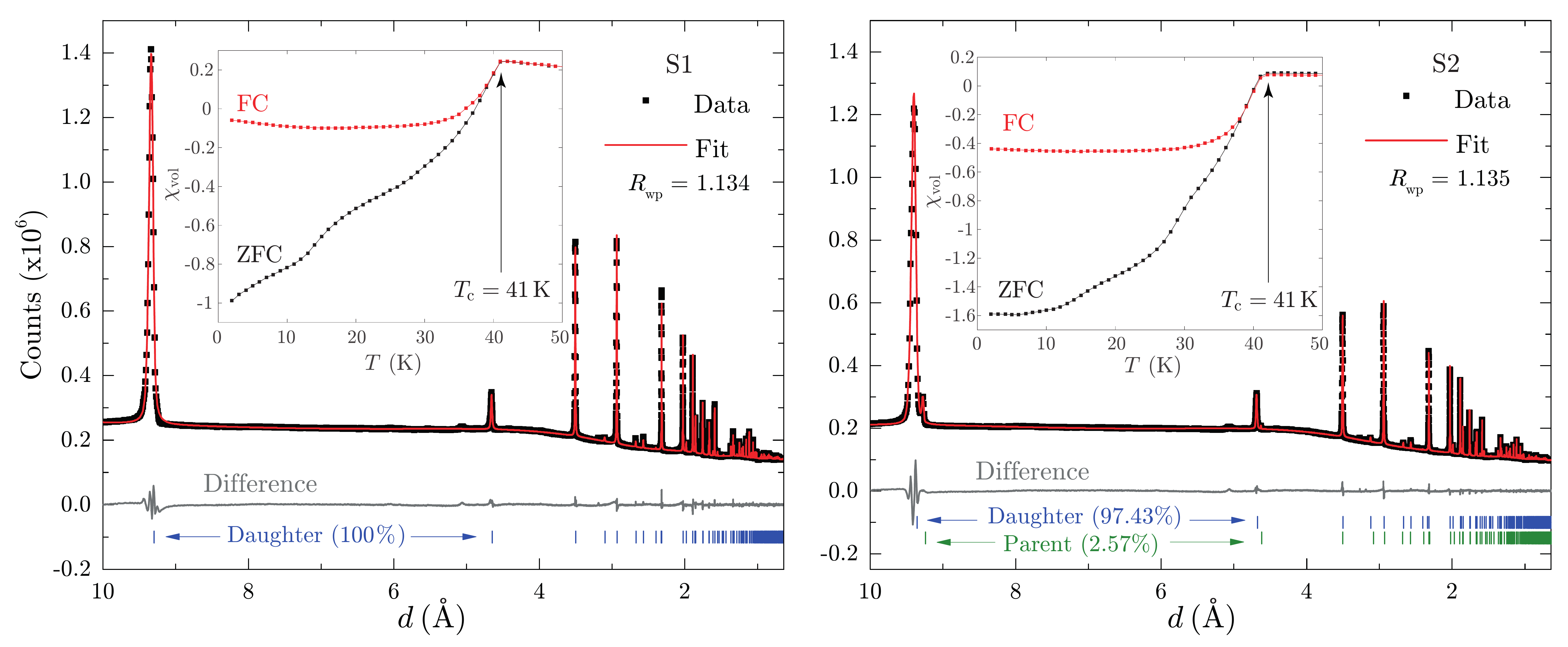}
	\caption{Main Panels: X-ray powder diffraction data measured on beamline I11 at the Diamond Light Source for two different deuterated powder samples S1 (left) and S2 (right) of Li$_{1-x}$Fe$_{x}$ODFe$_{1-y}$Se after lithiation. Neutron scattering data presented in later figures were recorded from the combined sample S1 + S2. Black points are experimental data and red lines are a fit obtained by Rietveld refinement with space group $P4/nmm$ and tetragonal lattice parameters $a = 3.78088(1)$\,\AA, $c = 9.30020(7)$\,\AA~(Daughter S1 = Li$_{0.84}$Fe$_{0.16}$ODFe$_{0.99}$Se) and $a = 3.77538(1)$\,\AA, $c = 9.36136(7)$\,\AA~(Daughter S2 = Li$_{0.83}$Fe$_{0.17}$ODFe$_{0.97}$Se). The sample in S2 is found to have a small impurity of the unlithiated parent sample (parent composition = Li$_{0.81}$Fe$_{0.19}$ODFe$_{0.94}$Se). Insets: Measured dimensionless volume magnetic susceptibility $\chi_{\rm vol}$ on field-cooling (FC) and zero-field-cooling (ZFC) for the same two samples. $T_{\rm c}$ is determined using the first data point at which $\chi_{\rm vol}$ begins to reduce with cooling. No correction has been made for demagnetisation.}
	\label{Characterisation_Data_Fig}
\end{figure*}

Several competing theoretical models have been proposed to explain the superconductivity and remarkably high $T_{\rm c}$ in iron-based superconductors whose Fermi surface consists only of electron pockets, such as Li$_{1-x}$Fe$_{x}$OHFe$_{1-y}$Se. A variety of novel pairing mechanisms based on magnetic \cite{Guterding2015, Wang2011c, Saito2011, Korshunov2008, Hao2014, Pandey2013} and orbital \cite{Ong2016, Nica2015} fluctuations have been suggested, with the theories predicting different superconducting gap symmetries, including sign-preserving s$^{++}$-wave and sign-changing s$^{+-}$ or d-wave. A conclusive determination of the gap symmetry and sign distribution on the Fermi surface, which is crucial for distinguishing between these theories, has so far proved elusive.


The focus of the present work is on magnetic fluctuations in Li$_{1-x}$Fe$_{x}$ODFe$_{1-y}$Se, the deuterated form of the Li$_{1-x}$Fe$_{x}$OHFe$_{1-y}$Se superconductor, which we have measured using neutron spectroscopy. Many experimental studies have found that magnetic correlations couple to superconductivity in the iron-based superconductors \cite{Lumsden2010, Dai2012, Dai2015, Inosov2016}, and the results have provided a strong stimulus for theories that invoke magnetic fluctuations as a key ingredient in the unconventional superconducting pairing interaction \cite{Mazin2008,Scalapino2012,Chubukov2012}.

One piece of evidence used in support of magnetic pairing is the spin resonance peak, which has been observed widely in the iron arsenide and selenide superconductors \cite{Johnston2010,Stewart2011,Lumsden2010,Dai2012}. The spin resonance is a collective spin excitation that appears below $T_{\rm c}$ and whose signature is a peak in the neutron scattering spectrum. The resonance peak is centered on a characteristic wave vector ${\bf Q}_{\rm res}$ which is often close to or the same as the propagation vector of an antiferromagnetic phase that borders superconductivity, and the peak appears in a narrow range of energy $E_{\rm res} \simeq 5$--$6 k_{\rm B}T_{\rm c}$ just below the maximum of the superconducting gap.  In weak coupling spin fluctuation theories of superconductivity the spin resonance is caused by strong scattering between points on the Fermi surface that are connected by ${\bf Q}_{\rm res}$  and have opposite signs for the superconducting gap function. The resonance tends to be sharp in $\bf Q$ when the relevant parts of the Fermi surface are well nested \cite{Maier2008,Korshunov2008,Mazin2009,Hirschfeld2016}. Measurements of the resonance peak, therefore, can provide very useful information on the symmetry of the superconducting gap and the underlying band structure.

Our study provides evidence for a spin resonance in Li$_{1-x}$Fe$_{x}$ODFe$_{1-y}$Se ($T_{\rm c}$ = 41\,K) at a wave vector ${\bf Q}_{\rm res}$ of approximately $(\pi, 0.67\pi)$, which is remarkably similar to that found in measurements on a sample of Li$_{x}$(ND$_{3}$)$_{1-y}$(ND$_{2}$)$_{y}$Fe$_{2}$Se$_{2}$ with the same $T_{\rm c}$ \cite{Taylor2013}, and close to $(\pi, \pi/2)$ as found in FeSe intercalated with alkali metal ions \cite{Park2011,Friemel2012a,Taylor2012,Friemel2012b,Wang2016}. This wave vector can plausibly be explained by nesting between the electron Fermi surface pockets measured by ARPES on single crystals with the same electron doping level as our sample \cite{Niu2015, Zhao2016}, thereby ruling out a sign-preserving s$^{++}$ pairing symmetry in Li$_{1-x}$Fe$_{x}$ODFe$_{1-y}$Se.

\section{Experimental details}

Two separate batches of polycrystalline Li$_{1-x}$Fe$_{x}$ODFe$_{1-y}$Se, with masses of 9.18\,g (S1) and 8.17\,g (S2), respectively, were synthesized via the novel lithiation method detailed in Ref.~\cite{Sun2015}. Fully-deuterated samples were prepared in order to reduce the incoherent background in the neutron scattering measurements. All sample handling was performed in an inert atmosphere.  Synchrotron X-ray powder diffraction patterns taken on the Diamond Light Source I11 beamline \cite{Thompson2009} (Fig.~\ref{Characterisation_Data_Fig} main panels) show both to be high quality with almost no Fe vacancies in the FeSe plane and no detectable impurities except a small amount of unlithiated parent material in one of the samples. The refined compositions were Li$_{0.84}$Fe$_{0.16}$ODFe$_{0.99}$Se (S1) and Li$_{0.83}$Fe$_{0.17}$ODFe$_{0.97}$Se (S2) with a 2.6\% impurity of Li$_{0.81}$Fe$_{0.19}$ODFe$_{0.94}$Se in S2. Field-cooled and zero-field-cooled magnetic susceptibility data taken on a Superconducting Quantum Interference Device (SQUID) magnetometer (Fig.~\ref{Characterisation_Data_Fig} insets) show a high superconducting volume fraction and $T_{\rm c} \simeq 41$\,K in both samples. Evidence from a previous study on samples synthesised via the same method indicates that the impurity composition will either not superconduct or will do so only with a low $T_{\rm c}$ ($< 10$\,K) \cite{Sun2015}, and in light of this observation and its low mass fraction in the sample we expect the impurity to produce no measurable effect on the neutron spin resonance measurements discussed later in this work. We note that there is no essential difference in susceptibility between the deuterated samples measured here and published data from equivalent samples containing natural hydrogen \cite{Sun2015}, confirming that deuteration has no effect on the bulk superconducting properties of this system. Magnetic susceptibility and X-ray diffraction measurements confirmed that the samples remained unchanged after the experiment.


Inelastic neutron scattering measurements were performed on the Merlin time-of-flight chopper spectrometer at
the ISIS Facility \cite{Bewley2006}. The two powder samples S1 and S2 were sealed in separate aluminium foil packets in concentric annular geometry inside a cylindrical aluminium can with diameter 4\,cm and height 4\,cm . The can was mounted inside a closed-cycle cryostat.
Spectra presented here were taken at a range of temperatures between 6.5 and 62\,K with neutron incident energy $E_{\rm i} = 80$\,meV and normalised to the scattering from a standard vanadium sample to place the data on an absolute intensity scale of mb\,sr$^{-1}$\,meV$^{-1}$\,f.u.$^{-1}$, where 1\,mb = $10^{-31}$\,m$^{2}$ and f.u. stands for one formula unit of Li$_{1-x}$Fe$_{x}$ODFe$_{1-y}$Se (such that all scattering in this work is presented per Fe site in the FeSe layer), although we note that we have not corrected for the strong neutron attenuation of the sample. We estimate from the FWHM of the elastic line that the energy resolution in this configuration is 5\,meV at the elastic line and 3.7\,meV at an energy transfer of 24\,meV. All spectra presented here have also been corrected for the Bose population factor $\left\{1-\exp(-E/k_{\rm B}T)\right\}^{-1}$, where $E$ is the neutron energy transfer and $T$ is the temperature, so that the presented quantity is the dynamical susceptibility $\chi''(Q,E)$.

\section{Results}

\begin{figure}
	\centering
	\includegraphics[width=0.9\columnwidth]{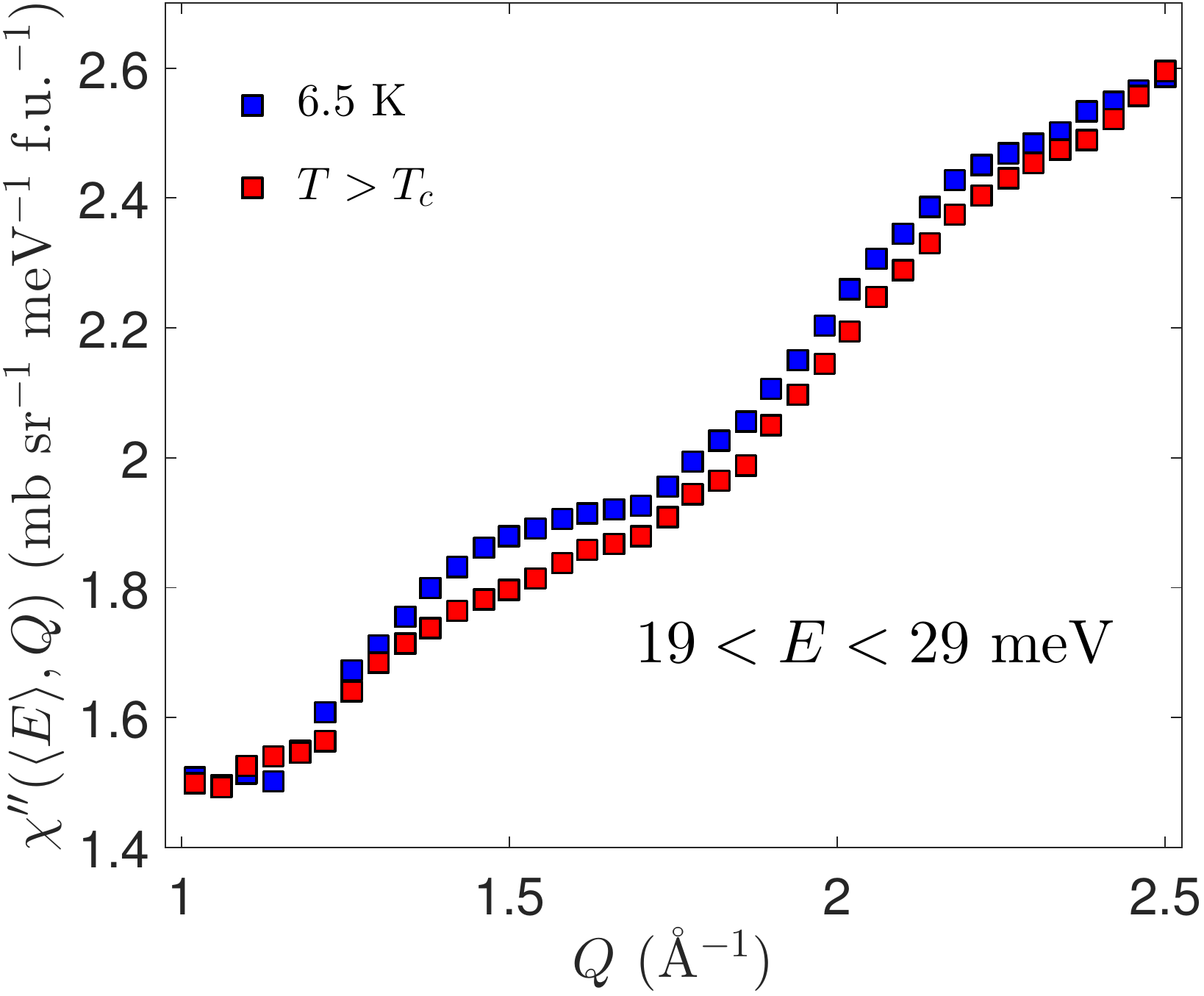}
	\caption{Measured dynamical susceptibility as a function of momentum transfer $Q = |{\bf Q}|$ averaged over the range of energy transfer $19 < E < 29$\,meV. This quantity was obtained by measuring the inelastic neutron scattering from a polycrystalline sample with average composition Li$_{0.84}$Fe$_{0.16}$ODFe$_{0.98}$Se at $T = 6.5$\,K and $T > T_{\rm c}$ after normalisation to the bose population factor $\left\{1-\exp(-E/k_{\rm B}T)\right\}^{-1}$. Data for $T > T_{\rm c}$ are an average of runs recorded at 51\,K and 62\,K. The formula unit (f.u.) used for normalisation is that of Li$_{0.84}$Fe$_{0.16}$ODFe$_{0.98}$Se.}
	\label{Raw_Data_Fig}
\end{figure}

Figure~\ref{Raw_Data_Fig} presents scattering as a function of momentum transfer $Q$ averaged over the energy transfer range $19 < E < 29$\,meV. The two sets of measurements shown are for temperatures of $T=6.5$\,K and $T > T_{\rm c}$, where the $T > T_{\rm c}$ curve is an average of data collected at 51\,K and 62\,K to improve the statistics. The justification for averaging is that the 51\,K and 62\,K data show no observable difference over the $(Q, E)$ range of interest after correction for the Bose factor. The general increase in scattering with $Q$ is due to scattering from phonons, but the $T=6.5$\,K curve has a clear enhancement in spectral weight over the $T > T_{\rm c}$ curve in a broad range of $Q$ from around 1.2 to 2.6~\AA$^{-1}$. We attribute the excess scattering at low temperatures to the spin resonance appearing in the superconducting state, since any change in the phonon background is taken into account by the Bose population factor correction.

\begin{figure}
	\centering
	\includegraphics[width=0.9\columnwidth]{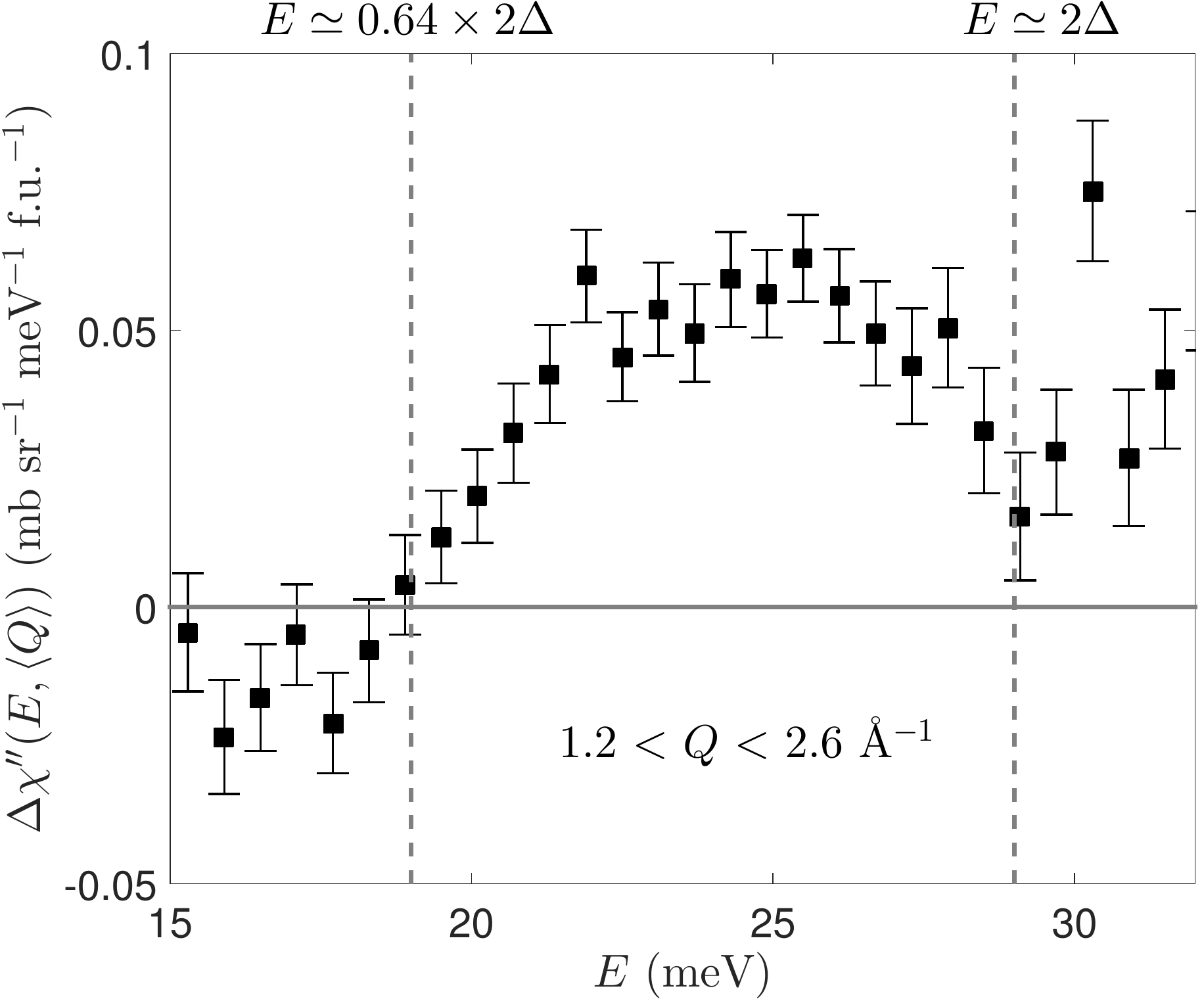}
	\caption{Excess neutron scattering intensity $\Delta \chi''(E, \langle Q \rangle) =  \chi''(E, \langle Q \rangle, T = 6.5\,{\rm K}) - \chi''(E, \langle Q \rangle, T > T_{\rm c})$ , after normalisation to the Bose population factor. The normal state intensity is the average of runs recorded at 51\,K and 62\,K as in Fig.~\ref{Raw_Data_Fig} and the signal is averaged across the full double peak structure, i.e. $1.2 < Q < 2.6$~\AA$^{-1}$. Dashed vertical lines mark $E \simeq 2\Delta$ and $E \simeq 0.64\times 2\Delta$, where $\Delta \simeq 14.5$\,meV is the superconducting gap from ARPES \cite{Zhao2016}.}
	\label{Energy_Cut_Fig}
\end{figure}

Further evidence that this is the spin resonance can be seen in the energy dependence of excess scattering at 6.5\,K relative to $T > T_{\rm c}$ (Fig.~\ref{Energy_Cut_Fig}), which shows a broad hump between about 20 and 30\,meV. Using $T_{\rm c} = 41$\,K and the gap energy $\Delta \simeq 14.5$\,meV measured by ARPES on samples of this material with the same electron doping level and $T_{\rm c}$ \cite{Zhao2016}, we see that this signal is fully consistent with other unconventional superconductors where the resonance appears at $E_{\rm res} \simeq 0.64\times2\Delta\ ( \simeq 19$\,meV), or more approximately at $E_{\rm res} \simeq 5.8k_{\rm B}T_{\rm c}\ ( = 21$\,meV) \cite{Yu2009}.

\begin{figure}
	\centering
	\includegraphics[width=0.9\columnwidth]{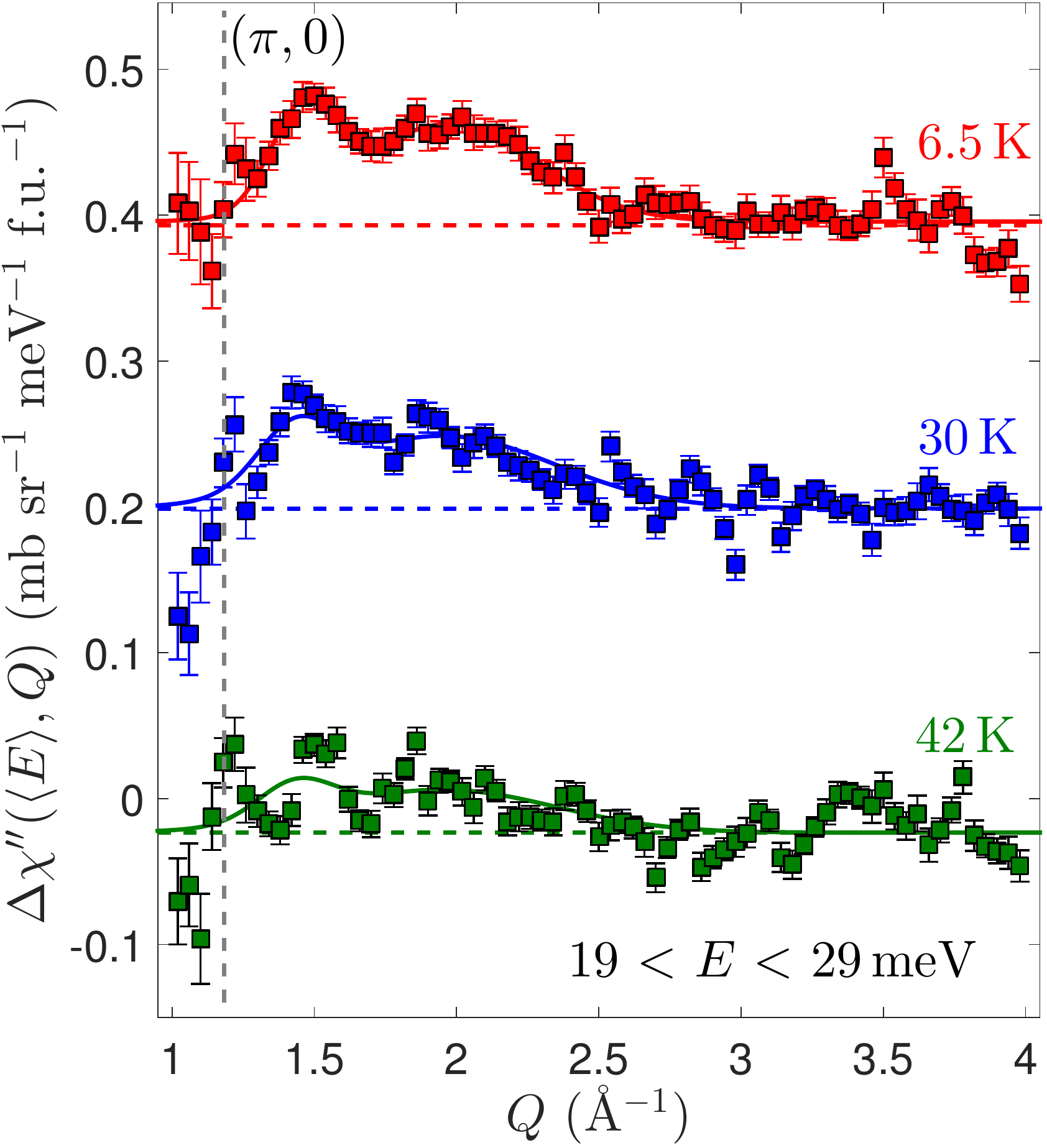}
	\caption{Excess neutron scattering intensity as a function of $Q$ at 6.5\,K, 30\,K and 42\,K relative to the average of runs recorded at 51\,K and 62\,K, as in Fig.~\ref{Raw_Data_Fig}. The 30\,K and 6.5\,K data have been offset vertically by 0.2 and 0.4 units, respectively. The double peak structure observed at low $Q$ is attributed to the spin resonance. Dark coloured lines are a best fit to two Gaussians plus a constant background at each temperature, as discussed in the main text, while horizontal broken lines are the constant background at each temperature intended as a guide to the eye. The dashed vertical line represents the $Q$ position corresponding to the 2D wave vector $(\pi, 0)$ in the 1-Fe square lattice Brillouin Zone.}
	\label{Q_Cut_Fig}
\end{figure}

\begin{table*}
	\caption{Best-fit parameters for a two-Gaussian lineshape plus a constant background obtained from a least-squares fit to the 6.5\,K data in Figure~\ref{Q_Cut_Fig} (Li$_{0.84}$Fe$_{0.16}$ODFe$_{0.98}$Se) and the 5\,K data in Figure~\ref{ND3_Comparison_Fig} (Li$_{0.6}$(ND$_{2}$)$_{0.2}$(ND$_{3}$)$_{0.8}$Fe$_{2}$Se$_{2}$, from Ref.~\cite{Taylor2013}). The $Q_{i}$s are peak centres and $\sigma_{i}$s the corresponding standard deviations where $\sigma = {\rm FWHM}/2\sqrt{2 {\rm ln} 2}$. \\}
	\label{Fitting_Params_Table}
	\begin{tabular}{C{4.5cm}|C{1.5cm}C{1.5cm}C{1.5cm}C{1.5cm}}
		\hline
		\hline
		\\[-0.23cm]
		Composition & $Q_{1}$ (\AA$^{-1}$)& $\sigma_{1}$ (\AA$^{-1}$)& $Q_{2}$ (\AA$^{-1}$) & $\sigma_{2}$ (\AA$^{-1}$) \tabularnewline[3pt]
		\hline\\[-0.25cm]
		Li$_{0.84}$Fe$_{0.16}$ODFe$_{0.98}$Se &1.46(3) &0.12(4) &1.97(7) &0.32(8) \tabularnewline[2pt]
		Li$_{0.6}$(ND$_{2}$)$_{0.2}$(ND$_{3}$)$_{0.8}$Fe$_{2}$Se$_{2}$ & 1.43(6)& 0.27(7)& 2.07(7)& 0.21(8) \tabularnewline[2pt]
		\hline
		\hline
	\end{tabular}
\end{table*}

\begin{figure}
	\centering
	\includegraphics[width=0.8\columnwidth]{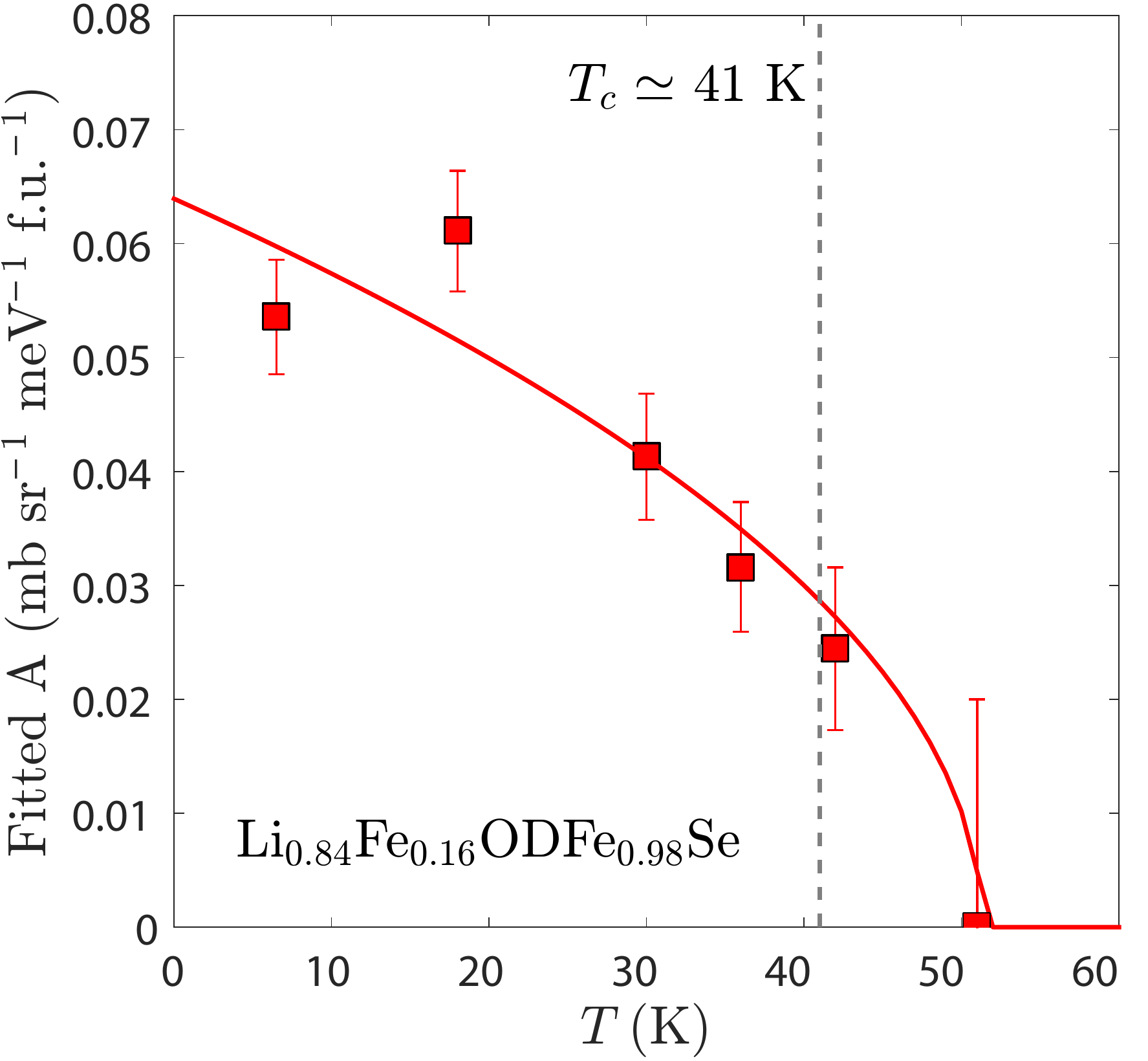}
	\caption{Temperature dependence of the fitted overall amplitude $A$ of the double-peak structure observed in Fig.~\ref{Q_Cut_Fig}. The line is the best fit to an order-parameter-like behaviour $I \sim (T-T_{0})^{1/2}$.}
	\label{Order_Param_Fig}
\end{figure}

A plot of the excess scattering as a function of $Q$ averaged over the energy range from 19\,meV to 29\,meV is given in Fig.~\ref{Q_Cut_Fig}. These energies correspond to the whole energy range for scattering below the pair-breaking energy from $0.64\times2\Delta \simeq 19$\,meV up to $2\Delta \simeq 29$\,meV. The excess scattering has a double peak structure at 6.5\,K with no clear change in the shape or position of either peak at intermediate temperatures in the superconducting state, for example 30\,K. Some other features are visible in the data, especially in the 30\,K and 42\,K curves, which though small are not accounted for by statistical noise. These could be the result of subtle variations in the background signal with temperature, e.g.~small changes in phonon modes, whose effect is magnified by the subtraction of two very similar signals.

In order to quantify the peak shapes and temperature dependence, we performed a fit similar to that in reference~\cite{Taylor2013} of the subtracted data at 6.5\,K to two Gaussian peaks plus a constant background, with peak heights $a_{i}$, centres $Q_{i}$ and widths $\sigma_{i}$ as well as the background all allowed to vary independently, yielding the fitted parameters shown in Table~\ref{Fitting_Params_Table}. For intermediate temperatures, the peak widths and centres were fixed to the values in Table~\ref{Fitting_Params_Table} and the ratio between the peak heights $a_{1}/a_{2}$ to its value at 6.5\,K such that only two parameters, the overall amplitude of the whole feature $A$ and the background, were refined. Within this treatment, $A$ shows a general increasing trend below $T_{\rm c}$ (Fig.~\ref{Order_Param_Fig}) which could plausibly be consistent with an order-parameter-like temperature dependence, $I \sim (T-T_{0})^{1/2}$. However, there is insufficient data to establish such a relationship conclusively.


\begin{figure}
	\centering
	\includegraphics[width=0.9\columnwidth]{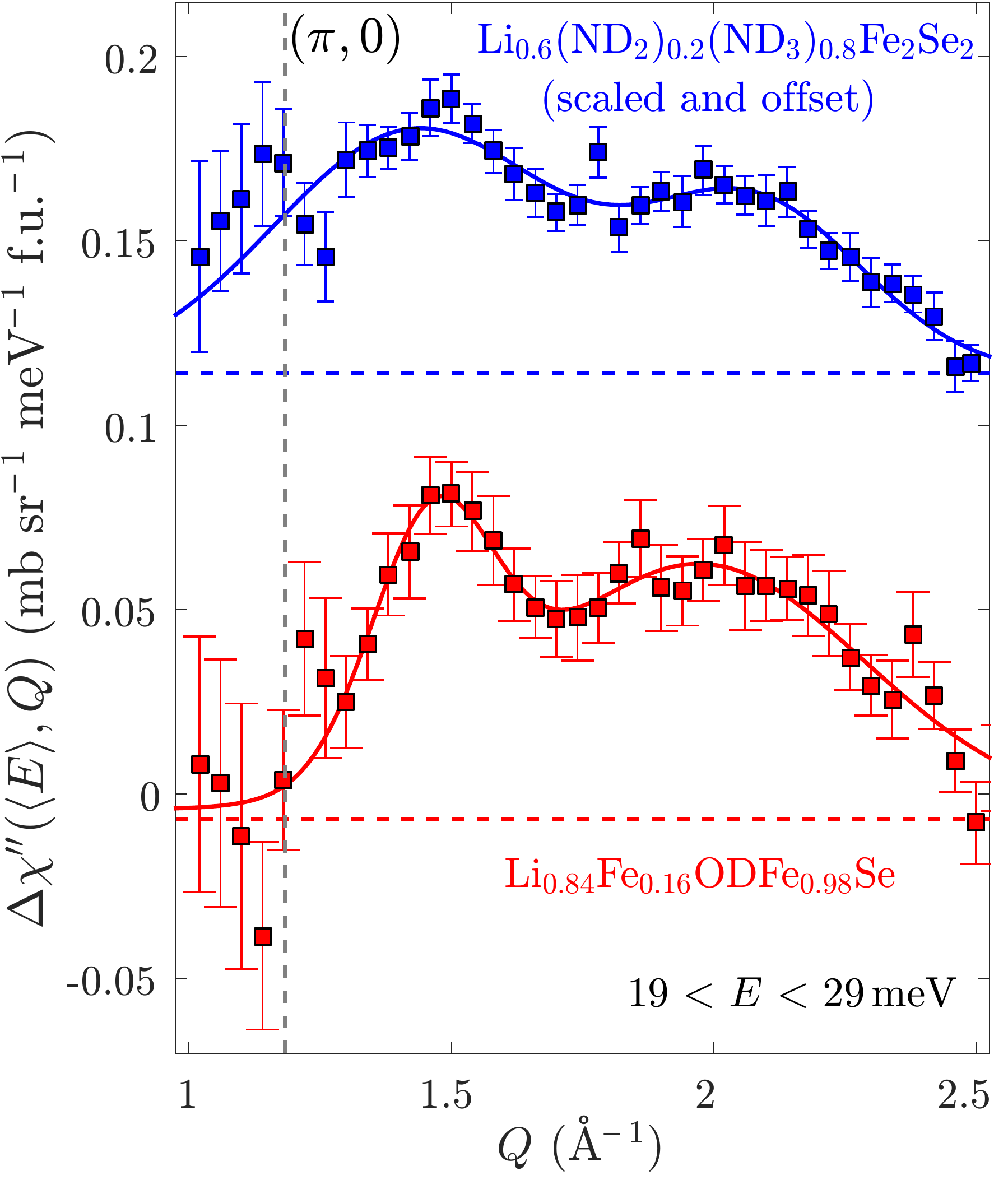}
	\caption{The 6.5\,K dataset from Fig.~\ref{Q_Cut_Fig} for Li$_{0.84}$Fe$_{0.16}$ODFe$_{0.98}$Se plotted with an equivalent dataset at 5\,K for the lithium ammonia/amide intercalated Li$_{0.6}$(ND$_{2}$)$_{0.2}$(ND$_{3}$)$_{0.8}$Fe$_{2}$Se$_{2}$ from Ref.~\cite{Taylor2013}. For ease of comparison, the Li$_{0.6}$(ND$_{2}$)$_{0.2}$(ND$_{3}$)$_{0.8}$Fe$_{2}$Se$_{2}$ dataset has been offset vertically and scaled by a constant factor so that the lower $Q$ peak maximum is the same height above background as that in the Li$_{0.84}$Fe$_{0.16}$ODFe$_{0.98}$Se data to take into account the differing attenuation between the two samples.}
	\label{ND3_Comparison_Fig}
\end{figure}

The spin resonance signal we observed here for Li$_{1-x}$Fe$_{x}$ODFe$_{1-y}$Se ($T_{\rm c} \simeq 41$\,K) is remarkably similar to that measured under very similar conditions on a powder sample of Li$_{0.6}$(ND$_{2}$)$_{0.2}$(ND$_{3}$)$_{0.8}$Fe$_{2}$Se$_{2}$ ($T_{\rm c} \simeq 43$\,K) \cite{Taylor2013}. Cuts along the $Q$ axis are presented for both materials in Fig.~\ref{ND3_Comparison_Fig} averaged over the same energy range. The excess scatting at low temperatures shows the same double-peak structure for each material, with peak positions at the same $Q$ to within experimental error. Fitted parameters are compared in Table~\ref{Fitting_Params_Table}. The most significant difference between the two datasets is a constant scale factor resulting in $\sim$ 3 times more scattering per Fe site for the ammonia intercalate. This scale factor applies to the whole spectrum including the elastic line and is explained by increased attenuation (neutron absorption and elastic scattering) due partly to the greater presence of Li in the LiOD-intercalate and partly to difference in the average neutron path lengths through the two samples. Another difference is some extra scattering from the ammonia-intercalate at low $Q \simeq 1.2$\,\AA$^{-1}$ which has the effect of widening the lower $Q$ peak and is not present for the LiOD-intercalate.

In order to gain information about the superconductivity from the data presented here it is necessary to know where the resonance occurs in the Brillouin zone. Using similar arguments to those presented in Ref.~\cite{Taylor2013} for the ammonia-intercalate which attribute the two peaks to scattering from resonance positions in the first and second Brillouin zones, it is also clear for the LiOD-intercalate that the resonance does not occur at the Fe square lattice wave vector $(\pi, 0)$, as found in pure FeSe \cite{Rahn2015} (c.f. vertical dashed lines in Figs.~\ref{Q_Cut_Fig} and \ref{ND3_Comparison_Fig}). Instead, it is found to be in the vicinity of $(\pi, \pi/2)$ and equivalent positions as seen in other FeSe intercalates, and is in fact best described by $(\pi, 0.67\pi)$. This is remarkably close to $(\pi, 0.625\pi)$ as predicted in one calculation~\cite{Maier2011}.

In Ref.~\cite{Taylor2013} it is suggested that the extra weight around $Q \simeq 1.2$\,\AA$^{-1}$ may be from a small secondary superconducting phase with a $(\pi, 0)$ resonance appearing around 10\,K, possibly due to some degree of phase separation or sample inhomogeneity. The sharper first order peak and complete lack of excess scattering at $Q$ corresponding to $(\pi, 0)$ in the data presented here indicates there is no such minority phase or sample inhomogeneity in the LiOD-intercalate.

Similarly, the initial studies of the LiOH-intercalate reported evidence for either antiferromagnetic (Ref.~\cite{Lu2015}) or ferromagnetic (Ref.~\cite{Pachmayr2015}) order. The low energy spin-wave excitations of any long-range magnetic order would have a characteristic scattering intensity at low energies which would disperse away from the respective ordering wave vectors. We did not observe any such scattering in the neutron spectrum of Li$_{0.84}$Fe$_{0.16}$ODFe$_{0.98}$Se, consistent with the lack of any magnetic Bragg peaks in neutron diffraction \cite{Sun2015, Zhou2016, Lynn2015}. The measurements constrain any magnetic order in optimally superconducting Li$_{1-x}$Fe$_{x}$ODFe$_{1-y}$Se either to a minority phase, or to a uniform phase with extremely small magnetic moments.

\section{Discussion}
\begin{figure}
	\centering
	\includegraphics[width=0.9\columnwidth]{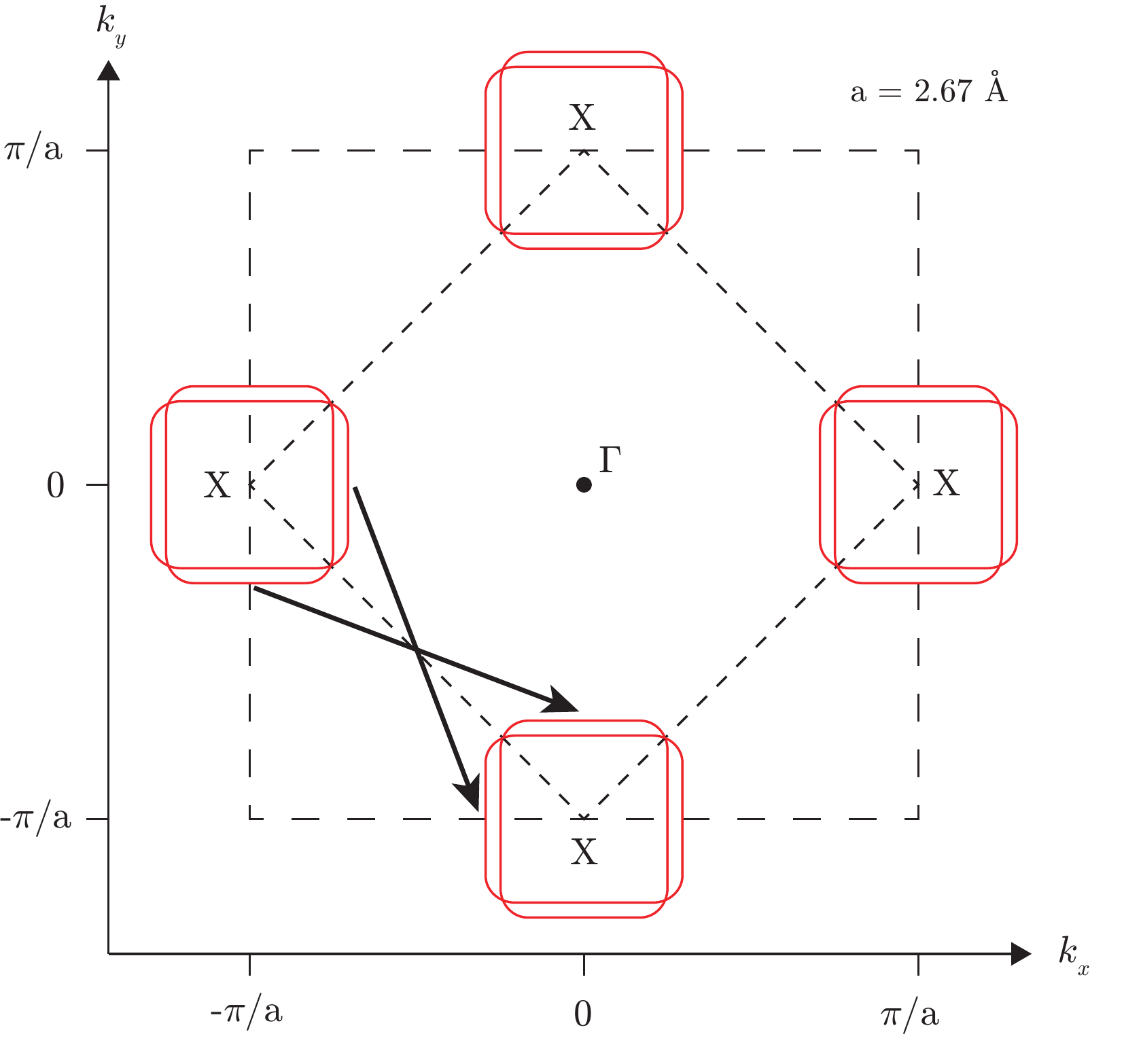}
	\caption{Schematic Fermi surface in the $k_{z} = 0$ plane for Li$_{1-x}$Fe$_{x}$OHFe$_{1-y}$Se, based on the calculation in Ref.~\cite{Maier2011} and qualitatively consistent with all current experiments. Brillouin zone boundaries are marked with thick dashed lines (1-Fe BZ) and thin dashed lines (2-Fe BZ) while the lattice parameter used here $a = 2.67$\,\AA~is that of the 1-Fe in-plane unit cell. $\Gamma$ and X labels mark special points of the 1-Fe Brillouin zone, with each X point surrounded by two box-like fermi sheets (red lines) which in reality are presumably hybridised at the points where they approach each other to avoid crossings. Thick arrows mark nesting vectors between straight, parallel sections of Fermi surface which cause a peak in nesting around $(\pi, 0.625\pi)$ as calculated in Ref.~\cite{Maier2011}. It should be noted that there is no experimental verification of the precise Fermi surface geometry at each X point due to limited resolution in ARPES, however the presented geometry is consistent with current experimental results.} 
	\label{FS_Fig}
\end{figure}

Recently, the band structure, Fermi surface and superconducting gap of LiOH-intercalated FeSe with the same level of electron doping and $T_{\rm c}$ as the sample used in this work were measured in two ARPES studies \cite{Niu2015, Zhao2016} and via scanning tunnelling spectroscopy \cite{Du2016}. Both ARPES studies found that the Fermi surface consists of almost identical, highly 2D electron pockets centred on $(\pi, 0)$ and $(0, \pi)$, i.e. the two X points of the 1-Fe square lattice Brillouin Zone, although theoretical calculations indicate that it is reasonable to expect two closely spaced Fermi sheets at each of these X points which are presumably not resolvable in ARPES and may hybridise with one another. The scanning tunnelling spectroscopy study provided strong evidence for this two-fermi-sheet scenario, showing that there are two different superconducting gaps. A schematic of this Fermi surface geometry is provided in Fig.~\ref{FS_Fig}, with the precise Fermi surface cross-sections represented by those calculated in Ref.~\cite{Maier2011}, although it should be noted that there is no experimental verification of these cross-sections due to limited resolution in ARPES.

Na{\"i}vely, nesting in this Fermi surface geometry would be expected to peak at $(\pi, \pi)$, i.e.~the vector connecting two X points at $(\pi, 0)$ and $(0, \pi)$, however it has been shown via theoretical calculations \cite{Maier2011, Pandey2013} that if the pockets have a non-circular cross section then it is possible for the nesting to have a broad plateau around $(\pi, \pi)$ and to reach its maximum at some other position. For example, a full calculation for the type of Fermi surface represented in Fig.~\ref{FS_Fig} shows the maximum is at $(\pi, 0.625\pi)$ \cite{Maier2011} due to enhanced nesting between parallel sections of the box-like Fermi surfaces as marked by bold arrows in Fig.~\ref{FS_Fig}. It is also possible to obtain a peak at $(\pi, \pi/2)$, as observed for the alkali-metal intercalates \cite{Park2011,Friemel2012a,Taylor2012,Friemel2012b,Wang2016}, or at $(\pi, 0.67\pi)$ as found in this work, by varying the precise Fermi surface geometry and size slightly. The detailed shape of the Fermi surface around the X points is not established in the ARPES data reported so far on Li$_{1-x}$Fe$_{x}$OHFe$_{1-y}$Se.

In weak-coupling spin fluctuation models the spin resonance is caused by nesting between sections of Fermi surface {\it with the opposite sign of the superconducting gap}, so our observation of a resonance at the very least rules out s$^{++}$ pairing in which the gap has the same sign at all places on all Fermi sheets.

There is evidence from scanning tunnelling spectroscopy \cite{Du2016} that the gaps on all Fermi sheets are nodeless and anisotropic, with the fitted gaps nevertheless preserving the 4-fold rotational symmetry of the crystallographic space group. If this is assumed to be correct then only three possible gap distributions remain, with either (i) the sign change between two decoupled Fermi sheets at the same X point, or (ii) the sign change  between equivalent Fermi sheets at different X points, or (iii) both (i) and (ii). In order to verify these assumptions a better experimental determination of the Fermi surface structure around a single X point is required, for example from higher resolution ARPES or quantum oscillations experiments.


\section{Conclusion}

In conclusion, we have observed a spin resonance appearing in the superconducting state of the FeSe intercalate Li$_{0.84}$Fe$_{0.16}$ODFe$_{0.98}$Se at a 2D wave vector close to $(\pi, \pi/2)$ as found in other FeSe-intercalates. Best agreement with the data is obtained with the wave vector $(\pi, 0.67\pi)$. We see no evidence for a $(\pi, 0)$ resonance, and the data are remarkably similar to previous measurements on the lithium ammonia/amide intercalate Li$_{0.6}$(ND$_{2}$)$_{0.2}$(ND$_{3}$)$_{0.8}$Fe$_{2}$Se$_{2}$ reported in Ref.~\cite{Taylor2013}. The observed wavevector $(\pi, 0.67\pi)$ is plausibly consistent with the nesting vector between pairs of 2D electron Fermi sheets around $(\pi, 0)$ and $(0, \pi)$ seen in ARPES and scanning tunnelling spectroscopy, which rules out conventional s$^{++}$ pairing. When considered in the light of evidence that the gap is nodeless, the observations constrain the sign of the gap to have one of three possible distributions on the Fermi surface.

{\it Note added.} Recently, an eprint appeared reporting neutron scattering measurements on single crystals of Li$_{1-x}$Fe$_{x}$ODFe$_{1-y}$Se \cite{Pan2016} which are consistent with the results described here.

\section{Acknowledgments}

This work was funded by the UK Engineering and Physical Sciences Research Council (EPSRC) (Grants EP/I017844 and EP/M020517/1) and the Leverhulme Trust (Grant RPG-2014-221). N.R.D. and M.C.R. are grateful for studentship support from the EPSRC and from the Clarendon Scholarship Fund of the University of Oxford, respectively. Work at Diamond Light Source was supported by award EE13284, and we thank Dr A. Baker and Dr C. Murray for assistance on beamline I11. We acknowledge useful discussions with C. V. Topping.

\bibliographystyle{apsrev4-1}
\bibliography{FeSe_Intercal_Bibliography}

\end{document}